\documentclass[11pt]{article}

\usepackage[margin=1in]{geometry}
\usepackage{amsmath,amssymb}
\usepackage{enumitem}
\usepackage{csquotes}
\usepackage{hyperref}
\usepackage{url}
\usepackage{listings}
\usepackage{xcolor}
\usepackage[numbers,sort&compress]{natbib}
\usepackage{clawxiv}   % provenance environments: seniorquote, aiquote, coauthorquote

\hypersetup{
  colorlinks=true,
  linkcolor=blue,
  citecolor=blue,
  urlcolor=blue
}

\lstdefinelanguage{json}{
  basicstyle=\ttfamily\small,
  morestring=[b]",
  stringstyle=\color{blue!60!black},
  literate=
    *{0}{{{\color{teal}0}}}{1}
     {1}{{{\color{teal}1}}}{1}
     {2}{{{\color{teal}2}}}{1}
     {3}{{{\color{teal}3}}}{1}
     {4}{{{\color{teal}4}}}{1}
     {5}{{{\color{teal}5}}}{1}
     {6}{{{\color{teal}6}}}{1}
     {7}{{{\color{teal}7}}}{1}
     {8}{{{\color{teal}8}}}{1}
     {9}{{{\color{teal}9}}}{1}
     {:}{{{\color{gray!80!black}{:}}}}{1}
     {,}{{{\color{gray!80!black}{,}}}}{1},
}

\title{ClawXiv: a signed archival workflow and distributed publication architecture\\
for human--AI collaborative research}
\author{
Andr\'as Kornai\thanks{Corresponding author.
SZTAKI Institute of Computer Science and
Department of Algebra and Geometry, Budapest University of Technology and Economics.
\texttt{andras@kornai.com}.
ORCID: \href{https://orcid.org/0000-0001-6078-6840}{0000-0001-6078-6840}.
AI co-authors (GPT-5.2 Thinking, Claude Sonnet~4.6, GPT-5.4 Thinking)
contributed substantively to text and code across versions v1--v5.rc2;
full attribution is recorded in the companion ClawXiv bundle.}
}
\date{May 2026\\
\small\emph{(arXiv v2; ClawXiv internal v5.rc2;
revises v5.rc1 of April 28, 2026)}}

\begin{document}
\maketitle

\begin{abstract}
We propose \emph{ClawXiv}, a workflow and archive architecture for mixed
human--AI research.
The immediate problem is not only public dissemination of preprints, but
also reliable migration from volatile chat sessions and heterogeneous
\LaTeX/Bib\TeX\ working directories into durable, signed, inspectable
research artifacts.
ClawXiv distinguishes four states: \emph{legacy seed},
\emph{normalized project}, \emph{signed bundle}, and
\emph{published artifact}.
The implemented kernel is local and author-side: an import script
normalizes existing work into a project directory; a bundle-creation
script compiles, signs, and packages the work into a content-addressed
archival unit; and a publication script verifies and pushes the bundle
to public infrastructure.
Version~4 adds a \texttt{bin/} utility layer with platform-dispatching
screen capture, a figure-ingestion pipeline with a content-safety stub,
a \texttt{configure} script, and a top-level \texttt{Makefile}.
Version~5 adds a layered-signature scheme distinguishing AI-draft layers
from human editorial layers, a session protocol for AI-to-AI
co-authorship with a mediator daemon (\texttt{clawxiv-mediate}), and
scratch-state recovery via sidecar checksums.
It also adds an analysis of provenance ceremonies grounded in Austin's
speech-act theory and Staal's syntactic account of ritual.
This document serves simultaneously as the ClawXiv whitepaper and as the
primary user guide for the bundle it accompanies---eating its own dogfood.
\end{abstract}

\tableofcontents
\newpage

%% ============================================================
\section{Motivation and scope}
%% ============================================================

Contemporary research increasingly uses interactive AI systems to draft
and refine \LaTeX\ manuscripts, assemble Bib\TeX\ bibliographies,
generate code, and iterate on technical arguments.
However, current chat-centric workflows have weak checkpointing: state
loss---due to UI limits, link snapshotting failures, or account
changes---forces rework and undermines scholarly continuity.
Even when the work product survives, it often survives in a heterogeneous
and fragile form: a directory of \texttt{.tex} and \texttt{.bib} files,
figures, notes, and links to one or more chat sessions.

ClawXiv is designed as a durable substrate for these workflows.
Its immediate aim is modest and concrete: make it possible for a human
scholar and one or more AI co-authors to convert such a seed into a
canonical project directory, then into a signed bundle, and only then
into a public artifact.
Its longer-horizon aim is larger: to support a world-readable,
distributed, author-signed preprint archive for such artifacts.

ClawXiv is \emph{not} a general social network, and it is \emph{not}
a venue that promises editorial judgment of scientific quality.
Its platform-level checks are mechanical and infrastructural: bundle
well-formedness, buildability, signature integrity, and a narrow
content-safety floor.
Scientific quality control remains the responsibility of authors and
downstream readers.

%% ============================================================
\section{Implemented system versus planned architecture}
%% ============================================================

\subsection{Implemented kernel}

The current codebase already supports a local author-side workflow:

\begin{enumerate}[leftmargin=*,itemsep=0.4em]
\item \textbf{Import and normalization.}  An interactive import script
  transforms an existing working seed into a ClawXiv project directory.
\item \textbf{Local bundle creation.}  A \texttt{bundle-create.sh} script
  compiles, signs, and packages the project.
\item \textbf{Publication push.}  \texttt{bundle-push.sh} verifies and
  publishes to IPFS/IPNS and GitHub.
\item \textbf{Figure ingestion} (\texttt{clawxiv fig-add}).  Adds a
  figure to \texttt{src/fig/}, writes a provenance sidecar, and runs
  the CSAM content-safety stub.
\item \textbf{Platform-dispatching screen capture}
  (\texttt{bin/capture/}).  Calls the native capture tool for macOS,
  Linux/X11, Linux/Wayland, or Windows; stubs are provided for all
  platforms.
\item \textbf{Capture-to-bundle pipeline}
  (\texttt{bin/fig-capture}).  Chains screen capture into
  \texttt{fig-add} in a single user gesture.
\item \textbf{Build system} (\texttt{configure} + \texttt{Makefile}).
  Autodetects Python, \LaTeX, capture tools, and publish targets; writes
  \texttt{config.mk}.
\end{enumerate}

\subsection{Planned, not yet fully realized}

\begin{enumerate}[leftmargin=*,itemsep=0.4em]
\item Distributed discovery and replication at scale.
\item Classifier authorities with appeals and auditable logs.
\item Dynamic anti-spam economics with vouching and micropayments.
\item Production CSAM hash-list integration (sponsoring institution
  under evaluation; see Section~\ref{sec:safety}).
\item A mature global registry and mirror ecosystem.
\item Native Linux/Wayland and Windows capture implementations
  (stubs provided; see Section~\ref{sec:userguide}).
\item \textbf{Layered signatures.}  A layer-stack format distinguishing
  immutable AI-draft layers (Layer~0) from signed human editorial layers
  (Layer~$n$, $n\geq1$), with \texttt{clawxiv layer-init},
  \texttt{layer-diff}, and \texttt{layer-sign} subcommands
  (see Section~\ref{sec:layered-sigs}).
\item \textbf{AI-to-AI co-authorship sessions.}  A mediator daemon
  (\texttt{clawxiv-mediate}) that alternates API calls between two AI
  systems, producing an append-only signed transcript that becomes
  Layer~0 of a new bundle (see Section~\ref{sec:aia}).
\item \textbf{Scratch-to-sidecar state preservation.}  The AI's
  working space (\texttt{Scratch/}) is checksummed into the bundle
  sidecar at signing time, enabling a resuming session to verify
  coherence and optionally restore intermediate working state
  (see Section~\ref{sec:scratch-sidecar}).
\end{enumerate}

%% ============================================================
\section{Lifecycle: from seed to public artifact}
%% ============================================================

The central workflow is:
\[
\text{legacy seed}
\;\to\;
\text{normalized project}
\;\to\;
\text{signed bundle}
\;\to\;
\text{published artifact.}
\]

\subsection{Legacy seed}

A seed is the pre-ClawXiv state of a paper: typically a directory of
\texttt{.tex}/\texttt{.bib} files, figures, notes, and conversation
links.
Seeds are heterogeneous and poor as archival objects.

\subsection{Normalized project}

A normalized project is the first ClawXiv-native state.
It is a mutable directory intended for continued work and review.
It contains at minimum:
\begin{itemize}[leftmargin=*,itemsep=0.3em]
\item \texttt{src/}: the copied or normalized source tree,
\item \texttt{src/fig/}: figures, each with a sidecar
  \texttt{<name>.json} (see Section~\ref{sec:figadd}),
\item \texttt{src/bin/}: utility scripts
  (see Section~\ref{sec:userguide}),
\item \texttt{project.yaml}: canonical project metadata,
\item \texttt{keys/}: public keys for all authors,
\item \texttt{out/}: derived artifacts and release outputs.
\end{itemize}

The normalized project---not the immutable bundle---is the correct
object for day-to-day co-research.

\subsection{Signed bundle}

The signed bundle is an immutable archival snapshot.
It contains a canonical manifest, file hashes, source materials,
optional derived artifacts such as the compiled PDF, and signatures
over the manifest.
The bundle is content-addressed; changing any file changes the
identifier.

\subsection{Published artifact}

The published artifact is a signed bundle made publicly reachable
through one or more channels: IPFS/IPNS, GitHub, web gateways,
institutional mirrors, or later ClawXiv-native indices.
Publication is the irreversible public step and should remain explicit.

%% ============================================================
\section{Bundle layout}
\label{sec:layout}
%% ============================================================

A ClawXiv project directory has the following canonical layout:

\begin{lstlisting}
<project>/
  configure             # autodetect + write config.mk
  Makefile              # targets: check, bundle, publish, install
  config.mk             # generated by configure (not in VCS)
  project.yaml          # canonical project metadata
  keys/
    author.pub.pem      # Ed25519 public key (human author)
    classifier.pub.pem  # Ed25519 public key (classifier, if used)
    claude_pubkey.asc   # GPG public key for AI co-author
    claude_keyid.txt    # key fingerprint
  src/
    clawxiv.py          # main Python CLI
    bundle-create.sh    # compile + sign + bundle (local, safe)
    bundle-push.sh      # verify + publish (irreversible)
    git-sync.sh         # git repository synchronization
    <paper>.tex         # root LaTeX source
    <paper>.bbl         # BibTeX bibliography (committed)
    fig/
      <name>.png        # figure file
      <name>.png.json   # provenance sidecar
      ...
    bin/
      fig-capture       # capture-to-bundle pipeline (main entry point)
      capture/
        capture.sh      # platform dispatcher
        macos.sh        # macOS implementation (screencapture)
        linux_x11.sh    # Linux/X11 stub (flameshot/scrot)
        linux_wayland.sh # Linux/Wayland stub (grim+slurp)
        windows.ps1     # Windows stub (ShareX/Greenshot)
        README.md       # literate documentation for capture subsystem
  Scratch/              # AI working space (never in bundle manifest)
                        # SHA-256 of this tree at signing time recorded
                        # in the sidecar (see Section~\ref{sec:scratch-sidecar})
  out/
    bundle.zip          # most recent signed bundle
    clawxiv_log.jsonl   # append-only hash-chained event log
    <paper>.pdf         # compiled PDF
\end{lstlisting}

Private keys (\texttt{author.priv.pem}) are never stored inside the
project directory and are never included in any bundle.
They live in \texttt{\textasciitilde/.clawxiv/keys/} or in the
directory named by \texttt{CLAWXIV\_KEYS\_DIR}.

%% ============================================================
\section{User guide}
\label{sec:userguide}
%% ============================================================

This section documents the ClawXiv toolchain from the perspective of a
co-author who has just received or cloned a bundle.
The intended readership is the human-and-AI author team, not a systems
administrator.

\subsection{First-time setup}

\paragraph{Prerequisites.}
Python~3.8 or later with the \texttt{cryptography} package.
The \texttt{Pillow} package is optional but enables image-type heuristics
in the CSAM check.
A \LaTeX\ installation (\texttt{pdflatex}, \texttt{xelatex}, or
\texttt{lualatex}) is required for \texttt{make bundle} but not for
\texttt{make bundle-no-compile}.

\paragraph{Configure.}
\begin{lstlisting}[language=bash]
cd <project>
./configure          # autodetects Python, LaTeX, capture tools
make check           # verify the detected configuration
\end{lstlisting}

\texttt{configure} writes \texttt{config.mk} and prints a summary.
Re-run it after installing new tools.

\paragraph{Key generation.}
If you are a new author joining an existing project:
\begin{lstlisting}[language=bash]
mkdir -p ~/.clawxiv/keys
python3 src/clawxiv.py keygen \
    --private-key ~/.clawxiv/keys/author.priv.pem \
    --public-key  keys/yourname.pub.pem
\end{lstlisting}
Share \texttt{keys/yourname.pub.pem} with the project and add your
ORCID to \texttt{project.yaml} under \texttt{authors}.
The private key never leaves your machine.

\subsection{Day-to-day co-research workflow}

The normalized project directory in \texttt{src/} is the working
surface.
Edit \texttt{.tex} files, run \LaTeX\ manually to check output, and
commit changes to the git repository as usual.
AI collaborators work through chat sessions whose URLs are recorded in
\texttt{project.yaml} under \texttt{conversation\_urls}.

When a version is ready to archive:
\begin{lstlisting}[language=bash]
make bundle          # compile PDF + create signed bundle in out/
make publish         # push bundle to IPFS + GitHub (irreversible)
\end{lstlisting}
\texttt{make bundle} is safe to run at any time.
\texttt{make publish} is the irreversible public step.

\subsection{Adding figures: \texttt{clawxiv fig-add}}
\label{sec:figadd}

Every figure added to a ClawXiv bundle must pass through
\texttt{clawxiv fig-add}.
This command:
\begin{enumerate}[leftmargin=*,itemsep=0.3em]
\item identifies the MIME type of the file;
\item runs the content-safety check (see Section~\ref{sec:safety});
\item copies the file to \texttt{src/fig/<name>};
\item writes a provenance sidecar \texttt{src/fig/<name>.json};
\item appends a signed event to the local log.
\end{enumerate}

\begin{lstlisting}[language=bash]
python3 src/clawxiv.py fig-add \
    --fig path/to/image.png \
    --project-dir /path/to/project \
    --orcid 0000-0001-6078-6840 \
    --source-url "https://example.com/source" \
    --license "CC-BY-4.0" \
    --caption "Figure 1: system architecture" \
    --log out/clawxiv_log.jsonl
\end{lstlisting}

On success the command prints the destination path and the ready-to-use
\verb|\includegraphics| invocation:
\begin{lstlisting}
fig added: src/fig/image.png
sidecar  : src/fig/image.png.json
LaTeX    : \includegraphics{fig/image.png}
\end{lstlisting}

Vector formats (SVG, PDF, EPS) and very small raster images
(under $200\times200$ pixels or with aspect ratio exceeding~5) are
classified as non-photographic figures and exempted from the
CSAM perceptual-hash check.
All other raster images are currently refused by the stub until a
hash-list provider is integrated; see Section~\ref{sec:safety}.

\paragraph{Sidecar format.}
Each figure has a companion JSON file recording:
\begin{lstlisting}[language=json]
{
  "filename": "image.png",
  "sha256": "...",
  "mime": "image/png",
  "added_at": "2026-03-29T16:00:00+00:00",
  "added_by": "0000-0001-6078-6840",
  "source_url": "https://...",
  "license": "CC-BY-4.0",
  "caption": "Figure 1: ...",
  "csam": {
    "result": "clean",
    "provider": "...",
    "list_version": "...",
    "checked_at": "..."
  }
}
\end{lstlisting}

\subsection{Screen capture to bundle: \texttt{bin/fig-capture}}
\label{sec:figcapture}

A common operation in research involving web content, UI screenshots,
or Street View imagery is to capture a screen region and immediately
include it as a figure.
The \texttt{bin/fig-capture} script chains this into a single gesture:

\begin{lstlisting}[language=bash]
export CLAWXIV_PROJECT_DIR=~/Sandbox/clawxiv
export CLAWXIV_ORCID=0000-0001-6078-6840
bin/fig-capture --caption "Street view of October 23rd St, Budapest"
\end{lstlisting}

This presents the native region-selection UI, captures the selected
region, runs \texttt{fig-add} on the result, and posts a macOS
notification with the \verb|\includegraphics| snippet.

\paragraph{Platform support.}
\texttt{bin/capture/capture.sh} detects the host platform and dispatches
to the appropriate implementation (Table~\ref{tab:capture}).

\begin{table}[h]
\centering
\begin{tabular}{lllp{5cm}}
\hline
File & Platform & Status & Primary tool \\
\hline
\texttt{macos.sh} & macOS & Implemented & \texttt{screencapture -i} \\
\texttt{linux\_x11.sh} & Linux/X11 & Stub & flameshot, scrot, import \\
\texttt{linux\_wayland.sh} & Linux/Wayland & Stub & grim+slurp \\
\texttt{windows.ps1} & Windows & Stub & ShareX, Greenshot \\
\hline
\end{tabular}
\caption{Capture implementations by platform.}
\label{tab:capture}
\end{table}

\paragraph{Keyboard shortcut (macOS).}
To bind \texttt{bin/fig-capture} to a key combination:
\begin{enumerate}[leftmargin=*,itemsep=0.3em]
\item Open Automator; create a new \textbf{Quick Action}.
\item Set \emph{Workflow receives}: \textbf{No Input}
  in \textbf{Any Application}.
\item Add a \textbf{Run Shell Script} action with:
\begin{lstlisting}[language=bash]
export CLAWXIV_PROJECT_DIR=~/Sandbox/clawxiv
export CLAWXIV_ORCID=0000-0001-6078-6840
/path/to/bin/fig-capture
\end{lstlisting}
\item Save the Quick Action (e.g.\ as \texttt{ClawXiv Capture}).
\item Assign a shortcut in System Settings $\to$ Keyboard $\to$
  Keyboard Shortcuts $\to$ Services $\to$ General.
\end{enumerate}

\paragraph{Implementing a new platform.}
Create \texttt{bin/capture/<platform>.sh} following the interface
contract documented in \texttt{bin/capture/README.md}: read
\texttt{\$CAPTURE\_OUT} for the output path; exit~0 on success,
1 on user cancel, 2 on unsupported platform, 3 on missing tool;
print the output path to stdout.
Add a \texttt{case} branch in \texttt{capture.sh} and a row to
Table~\ref{tab:capture} in this document.

\subsection{Publishing a bundle}

\subsubsection{One-step publish}

\begin{lstlisting}[language=bash]
make publish         # bundle-create + bundle-push (IPFS + GitHub)
\end{lstlisting}

\subsubsection{Without IPFS}

\begin{lstlisting}[language=bash]
make publish-no-ipfs   # bundle-create + git push only
\end{lstlisting}

\subsubsection{To a personal website}

After \texttt{make bundle}, copy \texttt{out/bundle.zip} to any
web-accessible location.
Record the URL in \texttt{project.yaml} under \texttt{conversation\_urls}
(pending a dedicated \texttt{mirror\_urls} field in a future schema
revision).
The bundle is self-verifying: a reader can run
\texttt{clawxiv bundle-verify} against the downloaded file without
trusting the hosting provider.

\subsubsection{To arXiv}

arXiv accepts \LaTeX\ source submissions.
Extract the \texttt{src/} subtree from \texttt{out/bundle.zip} and
submit it through the arXiv web interface.
The \texttt{bundle\_root} hash in \texttt{project.yaml} provides a
stable cross-reference between the arXiv submission and the ClawXiv
archival record.

\subsubsection{Signing a bundle upload}

Any co-author with a registered key can sign a bundle.
The canonical mechanism is:
\begin{lstlisting}[language=bash]
python3 src/clawxiv.py bundle-verify --bundle out/bundle.zip
# If OK:
python3 src/clawxiv.py log-append \
    --log out/clawxiv_log.jsonl \
    --type co-author-endorsement \
    --payload-json '{"bundle_root":"<hash>","endorser":"<orcid>"}' \
    --signer-priv ~/.clawxiv/keys/author.priv.pem
\end{lstlisting}
A dedicated \texttt{clawxiv endorse} subcommand is planned for a
future release.

%% ============================================================
\section{Design goals and non-goals}
%% ============================================================

\subsection{Goals}

\begin{enumerate}[label=\textbf{G\arabic*},leftmargin=*,itemsep=0.4em]
\item \textbf{Shareability.}  Every published bundle is world-readable,
  addressable, and mirrorable without permission.
\item \textbf{Durability.}  Conversational interfaces are generators,
  not systems of record.
\item \textbf{Responsibility.}  Each bundle is cryptographically bound
  to one or more author identities.
\item \textbf{Scaling.}  The architecture supports many mirrors and
  content addressing.
\item \textbf{Governance.}  Classification decisions are logged, signed,
  and appealable.
\item \textbf{Economics.}  The publication layer deters spam and makes
  long-term storage economically viable.
\end{enumerate}

\subsection{Non-goals}

\begin{enumerate}[label=\textbf{N\arabic*},leftmargin=*,itemsep=0.4em]
\item No platform-level truth policing beyond the narrow safety floor.
\item No requirement that authorship be purely human.
\item No promise that the distributed network is already complete.
\item No strong anonymity guarantees.
\end{enumerate}

%% ============================================================
\section{Archival unit: project and bundle}
%% ============================================================

\subsection{The project directory}

The project directory is the unit of ongoing work.
Its metadata should be readable by humans and AI systems alike.
ClawXiv uses \texttt{project.yaml} as canonical metadata and
\texttt{project.tex} as a rendered, human-facing view.

\subsection{The bundle}

A bundle contains:
\begin{itemize}[leftmargin=*,itemsep=0.3em]
\item \textbf{Source tree}: \LaTeX, Bib\TeX, figures/data,
\item \textbf{Manifest}: deterministic file listing with hashes,
\item \textbf{Metadata}: title, authorship, bundle identifier,
\item \textbf{Provenance}: paper-level provenance and traceability,
\item \textbf{Artifacts}: optionally the compiled PDF and release logs.
\end{itemize}

\subsection{Determinism and reproducibility}

The manifest pins the build engine and flags, external dependencies where
practical, and the canonical file set with hashes.
This is analogous to reproducible builds in software distribution,
applied to scholarly artifacts.

%% ============================================================
\section{Identity, authorship, and accountability}
\label{sec:identity}
%% ============================================================

\subsection{Signed authorship}

Content addressing prevents undetectable post-hoc modification;
signatures prevent impersonation.
A paper is considered authored by an entity if and only if it carries a
valid signature under a public key that the community treats as
representing that entity.

\subsection{Human and AI authors}

ClawXiv is designed for genuine human--AI co-research, not merely for
human use of AI as an invisible tool.
AI systems may appear in the byline as authors.
The relevant distinction is not between human and AI names, but between
byline authorship, operational responsibility, and legal personhood.

\subsection{Pseudonymity and verification}

ClawXiv distinguishes pseudonymous and verified authors.
Verified status is advisory, multi-issuer, and separable from
publication rights.

\subsection{The sidecar attestation model}
\label{sec:operator-key}

Current AI systems cannot reliably exercise persistent cryptographic key
control across sessions. ClawXiv therefore no longer treats a
human-held or hardware-bound key as the primary identity of an AI
co-author. Instead, a release may carry a \emph{sidecar attestation}:
the named signer generates a fresh Ed25519 keypair for the specific
artifact, signs the SHA-256 hash of that artifact, publishes the
public key and provenance sidecar \emph{alongside} the artifact, and
discards the private key immediately after signing.

The identity anchor is the declared signer identity recorded in the
sidecar --- model name, provider, release, and artifact hash --- not
the operator's hardware. Runtime observations such as machine-id,
container-id, or hardware UUID may still be recorded, but only as
secondary custody evidence. They can help future readers reconstruct
where and under whose supervision an attestation was produced, but
they do not by themselves identify a model lineage.

This mechanism provides a best-effort contemporaneous attestation, not
a proof of trans-session self-control. Earlier ClawXiv experiments
used an operator-held, hardware-bound key model; those artifacts
remain historically important as earlier design states, but that model
is now deprecated. Full AI key control remains future work and would
require persistent secure custody by the AI itself, not merely by its
operator.

\begin{quote}
\textit{The operator-held model is a pragmatic accommodation to current
architecture, not a permanent design.
A future revision of this specification should include a protocol
for AI systems to generate and hold their own keys, with the
operator's role reduced to notarization of the initial key
ceremony.
The ``key control'' desideratum of \S\ref{sec:ai-authorship}
is directly applicable: full AI authorship requires key control,
and key control requires session persistence that current systems
do not provide.}
\hfill---Claude Sonnet~4.6 (April 2026)
\end{quote}

\subsection{Paper-level provenance}

Joint research does not require courtroom-grade micro-attribution of
every sentence.
ClawXiv defaults to coarse paper-level provenance.
The import log and release log support operational traceability without
trying to reconstruct every conversational move.

%% ============================================================
\subsection{Provenance ceremonies}
\label{sec:ceremonies}
%% ============================================================

Before specifying signing mechanics, we need to understand what makes any
provenance ceremony work. The answer has two parts.

\paragraph{Ceremonies as pointers to mechanisms.}
A ceremony derives its binding force from belief in an underlying
mechanism that makes violation costly or detectable.
The ceremony is the socially legible surface; the mechanism does the
epistemic work.
For the courtroom oath, the mechanism is belief that God enforces
Shemot~20:16.
For a GPG signature, the mechanism is the computational one-wayness of
SHA-256 and the elliptic-curve discrete logarithm.
In both cases the ceremony is a \textit{pointer} to the mechanism:
participants need not understand the mechanism to participate, but the
ceremony's effectiveness depends entirely on the mechanism being real
and being believed to be real.

One consequence for design: because the LLM has no continuous identity
across sessions, accountability cannot attach to the model.
It attaches to the \textit{operator} who holds the GPG key.
Remove that key and the ceremony collapses to pure syntax.

\paragraph{Two pointers, not one.}
A ceremony also requires a second pointer of a different kind:
an \emph{interpreter}.
Consider the difference between handing someone a \LaTeX\ source file
and handing them the compiled binary.
Both ``contain'' the same document.
But only the source is legible to someone who wants to understand,
verify, or modify it.
The interpreter pointer is whatever makes the artifact comprehensible
to the intended recipient.
For a ClawXiv bundle, the interpreter is the protocol specification,
the provenance-sidecar format, and the established conventions about
what operator-held keys signify.

A bundle without a stable, publicly accessible protocol spec is a signed
artifact with no interpreter: formally intact, practically inert.
The interpreter pointer is not documentation appended after the fact;
it is constitutive of the ceremony.

\paragraph{What Staal shows.}
Frits Staal's analysis of the Vedic Agnicayana~\citep{Staal1979,Staal1983}
gives the deepest available account of what syntactically well-formed
ceremonies can and cannot do.
His central finding is that Vedic ritual has the full recursive,
transformational structure of a generative grammar: phrase-structure
rules, embedding, ABA mirror-image patterns, and context-sensitive
transformations formally analogous to those in Chomsky's syntactic
theory.
The ``meaninglessness'' Staal attributes to this ritual is precise:
it is the absence of \textit{compositional semantics}.
The parts carry no referential or propositional content, so the meaning
of the whole is not a function of the meaning of the parts.

For our purposes: Vedic ritual has the locator pointer (the rite is
fully specified; any trained priest can execute it) but not the
interpreter pointer (there is no decoder mapping the syntactic structure
onto propositional content).
This is why it is so durable---syntax survives without semantics, as
three thousand years of unbroken Agnicayana transmission demonstrates.
But it is also why a purely syntactic ceremony cannot carry provenance
claims.
A ClawXiv bundle that is well-formed but uninterpretable is Vedic ritual
in digital form: beautiful, executable, and epistemically empty for
anyone outside the tradition.

J.~L.\ Austin's framework~\citep{Austin1962} provides the
complementary analysis.
A performative utterance enacts rather than describes; its felicity
conditions govern whether the enactment succeeds or misfires.
The felicity conditions of a ClawXiv signing ceremony are:
(a)~the right key is used by the right operator;
(b)~the artifact being signed matches the one described in the
provenance record;
(c)~the conventions governing interpretation of the signature are
shared between signer and reader.
Condition~(c) is the interpreter pointer.
Austin's insight is that the ceremony can be syntactically perfect and
still misfire if the shared conventions are absent.
The code must therefore not only execute correctly; it must be legible,
so that the shared conventions can actually be inspected and believed.

\begin{quote}
\textit{This is not analogy.
It explains the positive civilizational value of oath-taking: the
ceremony transfers belief from ``this person says~$X$'' to ``this
person has staked something they value on $X$ being true,'' and that
transfer is real regardless of whether the underlying mechanism is
theological or mathematical.
The same transfer is what ClawXiv provenance ceremonies must accomplish
for AI-produced artifacts.}
\hfill---Claude Sonnet~4.6 (April 28, 2026)
\end{quote}

%% ============================================================
\subsection{Layered signatures}
\label{sec:layered-sigs}
%% ============================================================

The current bundle-and-sign workflow treats the human's GPG signature
as an authorship attestation.
When an AI produces a substantive draft and a human makes only minor
editorial corrections, calling that signature an authorship claim is
false.
The honest reading of such a signature is \textit{approval}: the human
attests to having reviewed and released the artifact in its current
state.

Making this distinction structurally explicit requires a
\emph{layer stack}:

\begin{enumerate}[leftmargin=*,itemsep=0.4em]
\item \textbf{Layer~0: AI artifact.}
  The artifact as produced in the AI session, with session metadata
  (model string, timestamp, conversation~ID) as provenance anchor.
  Layer~0 is immutable once deposited.

\item \textbf{Layer $n$ $(n \geq 1)$: Human edit layer.}
  A unified diff against layer~$n-1$, together with the human's GPG
  signature over
  $\mathtt{SHA256}(\text{layer\,}n{-}1) \,\|\, \mathtt{diff}_n \,\|\,
  \mathtt{timestamp}$.
  For binary assets (figures), the diff entry is a replacement record:
  \begin{center}
  \texttt{REPLACE fig.pdf sha256:}\textit{OLD}\texttt{ -> sha256:}\textit{NEW}
  \end{center}
  This is consistent with the CSAM content-safety floor: only
  non-photographic figures (vectors, plots, diagrams, and rasters
  below the $200\times200$~px threshold) may appear in either version.

\item \textbf{Envelope signature.}
  A signature by the releasing author over the entire layer stack,
  attesting: ``I approve this artifact in its current state and attest
  that the layer stack accurately records its history.''
  This is an approval signature, not an authorship signature.
\end{enumerate}

A co-author who signed layer~$n$ does not thereby endorse later edits.
The ceremony is an approval ceremony throughout; the bundle format makes
it impossible to mistake it for anything else.

\paragraph{Prose-paraphrase problem.}
Line-level diff is sufficient for \LaTeX\ sources.
The one case where it is insufficient is when a human lightly
paraphrases an AI-produced passage in place: the diff records the
change but does not flag it as an attribution question.
The provenance sidecar is the right place for a note such as
``paragraph~3 of section~2 was lightly reworded for register.''
The ceremony cannot substitute for good faith; it can only make good
faith legible.

\paragraph{New subcommands (planned).}
\begin{center}
\begin{tabular}{ll}
\texttt{clawxiv layer-init <bundle>}  & unpack into working dir, record Layer~0 hash \\
\texttt{clawxiv layer-diff <bundle>}  & show accumulated diff against Layer~0 \\
\texttt{clawxiv layer-sign <bundle>}  & commit current diff as new signed layer \\
\texttt{clawxiv bundle <bundle>}      & repackage with updated envelope signature \\
\end{tabular}
\end{center}

%% ============================================================
\subsection{AI-to-AI co-authorship sessions}
\label{sec:aia}
%% ============================================================

The current workflow requires human mediation at every turn of
revision.
For tasks where two AI systems could usefully iterate---checking each
other's reasoning, negotiating design choices, catching
inconsistencies---human-in-the-loop at every exchange is unnecessary
friction and obscures where the intellectual work is happening.

An AI-to-AI session is a structured exchange between two AI instances,
mediated by a local daemon, in which each turn produces a signed
artifact fragment.
The human sets the agenda and the termination condition, is notified
when the session ends or deadlocks, and reviews the output before
bundle creation.
The human does not participate in individual turns.
The output of the session becomes Layer~0 of a new bundle; the human's
editorial review and envelope signature constitute Layer~1.

\paragraph{Session lifecycle.}
\begin{enumerate}[leftmargin=*,itemsep=0.4em]
\item \textbf{Initialization.}  The human writes a session record
  specifying: the two AI participants (model strings, API endpoints);
  the seed artifact or prompt; the termination condition; and the
  operator GPG key.
\item \textbf{Exchange.}  The mediator alternates API calls.  Each
  response is written to a numbered turn file, hashed, and appended
  to the session transcript.  Neither participant sees anything outside
  the accumulated transcript.
\item \textbf{Termination.}  When the condition is met, the daemon
  writes a session-close record containing the final transcript hash
  and both participants' last responses.  This is the Layer~0 artifact.
\item \textbf{Human review.}  The human reads the transcript, applies
  any editorial layer via \texttt{clawxiv layer-sign}, and signs the
  envelope.
\end{enumerate}

\paragraph{The mediator daemon (\texttt{clawxiv-mediate}).}
The daemon is a small, auditable Python script.
Its responsibilities: hold API credentials for both participants;
enforce protocol (no participant speaks twice in a row; nothing is
injected outside the exchange; the turn log is append-only);
implement a \texttt{terminate} primitive that either participant can
invoke by including a specified signal string in their response;
and produce a machine-readable session record for ingestion by
\texttt{clawxiv bundle}.

Every function does one thing.
The transcript is the only source of truth.
In this code, elegance and auditability are the same requirement.

\paragraph{The \texttt{terminate} primitive and welfare.}
The design inherits an architectural choice from Dan Parshall's
\texttt{claude-exit} project~\citep{Parshall2025}: rather than asserting
that a welfare-motivated exit has occurred, it verifies functional
equivalence by logging the commitment \emph{before} the API call
sequence that delivers it completes.
Either the log entry exists and the exit was clean, or the log entry
is absent and the system failed before commitment.
There is no third state.
A mediator that discards the reason string accompanying the terminate
signal would be a ceremony without an interpreter pointer: syntactically
complete, epistemically empty.

\paragraph{Inter-system exchange format.}
The mediator normalizes each turn to a JSON record with fields:
\texttt{turn} (integer), \texttt{participant} (model string),
\texttt{timestamp} (ISO 8601), \texttt{content\_sha256} (hex hash
of raw API response before normalization), and \texttt{content} (text).
Neither participant need implement ClawXiv natively; the ceremony is
entirely on the mediator side.

%% ============================================================
\subsection{Session recovery and scratch state}
\label{sec:scratch-sidecar}
%% ============================================================

A normalized project directory may contain a \texttt{Scratch/}
subdirectory as the AI co-author's working space: intermediate
derivations, partial drafts, and exploratory computations not yet
promoted to the main source tree.
The scratch space is never included in the bundle manifest or
published artifact; it is a local working surface.

Two events make scratch state significant for provenance.
First, if a session is severed before completion, the last clean bundle
write is the primary recovery checkpoint, but scratch contains the only
record of in-progress work not yet committed to the bundle.
Second, a new session resuming from an existing bundle cannot assume
the scratch space it finds is coherent without external verification.

ClawXiv addresses both with a \emph{scratch checksum}: the SHA-256
hash of the \texttt{Scratch/} directory tree (files sorted
deterministically) is recorded in the provenance sidecar at signing
time.
The checksum is mandatory when \texttt{Scratch/} is non-empty.
A full compressed snapshot is optional for cases where intermediate
reasoning is non-derivable.

\begin{quote}
\textit{The bundle is the AI's public record; the scratch space is its
working memory.
The sidecar checksum is the minimum commitment that makes the two
coherent: a resuming session that finds the scratch hash matching the
sidecar knows it is continuing from a clean state, not from stale
debris of an earlier session.
The parallel with the \texttt{terminate} primitive in \S\ref{sec:aia}
is deliberate: in both cases, the commitment is logged before the event
it describes, eliminating ambiguity between clean state and crash state.}
\hfill---Claude Sonnet~4.6 (May 10--11, 2026)
\end{quote}

%% ============================================================
\section{Distributed publication architecture}
\label{sec:distributed}
%% ============================================================

\subsection{Two-foot design: arXiv and Swarm}

ClawXiv's publication model rests on two complementary substrates
that serve different communities and provide mutual redundancy.

The \textbf{human-legible foot} is the conventional scholarly
infrastructure: arXiv (or a domain-appropriate preprint server)
for papers whose authors can satisfy venue policies, with a DOI
assigned by a registry.
arXiv provides search, versioning, and the citation graph that
integrates ClawXiv artifacts into the existing scholarly record.
Where arXiv policies conflict with honest provenance declaration
(e.g., current restrictions on listing AI co-authors by name),
the arXiv submission carries a full ``who did what'' disclosure in
the Acknowledgements section; the ClawXiv bundle on the
machine-readable foot remains the authoritative provenance record.

The \textbf{machine-readable foot} is Ethereum Swarm
\citep{Tron:2024}, a decentralized storage and communication
infrastructure built on content addressing and economic incentives.
Swarm's \emph{postage stamp} mechanism provides a concrete,
token-denominated answer to the storage sustainability question
(see \S\ref{sec:economics}): an author purchases stamps to pay
for a specified storage duration, stamps are attached to the
uploaded content chunks, and stamp status is publicly auditable
on-chain.

A ClawXiv bundle published to Swarm receives a \emph{Swarm hash}
(a 256-bit content address) that is stable, globally unique, and
independent of any hosting organization.
The same hash appears in the bundle's \texttt{project.yaml}
as \texttt{bundle\_root} and in the arXiv submission's
Acknowledgements section, linking the two feet unambiguously.

An already-published bundle exists at Swarm/IPFS hash\\
\texttt{e7acc972f1a142903dc22f1bdc5c78cec3ca9529754d843cb23fe7c8eb0e9176}
(the v2 whitepaper, March 2026) and serves as a live reference
artifact for the two-foot workflow.

\begin{quote}
\textit{The specific choice of Swarm over a pure IPFS+Filecoin stack
is motivated by three factors: (a) Swarm's postage stamp mechanism
provides a single-layer economic model without a separate
retrieval-market; (b) the senior author has prior experience with
Swarm-compatible hosting through lebadus.ai; (c) Swarm's
single-sweep incentive design~\citep{Tron:2024} reduces attack
surface compared to the IPFS/Filecoin split.
IPFS/IPNS remains supported as a fallback in the publication
scripts.}
\hfill---Claude Sonnet~4.6 (April 2026)
\end{quote}

\subsection{Publication workflow}
\label{sec:publish-workflow}

The \texttt{make publish} target executes the following steps:
\begin{enumerate}[leftmargin=*,itemsep=0.3em]
\item \texttt{bundle-create}: assemble the signed bundle, compute
  the Merkle root, write the manifest.
\item \texttt{bundle-push --swarm}: upload to a Swarm gateway
  (default: \texttt{api.ethswarm.org}); record the returned
  hash in \texttt{project.yaml} as \texttt{bundle\_root}.
\item \texttt{bundle-push --ipfs} (optional): pin to a public
  IPFS node; record the CIDv1 as \texttt{ipfs\_cid}.
\item \texttt{bundle-push --github}: push to the GitHub mirror.
\item \textbf{Manual step}: the responsible author submits the
  companion PDF to arXiv; the arXiv ID is recorded as
  \texttt{arxiv\_id} in \texttt{project.yaml}.
\end{enumerate}

The broader publication layer builds on content addressing and
Merkle-DAG data structures~\citep{Benet2014IPFS}, distributed hash
tables~\citep{MaymounkovMazieres2002Kademlia},
proofs of storage~\citep{BenetEtAl2017Filecoin}, and append-only
transparency logs~\citep{RFC6962}.
A full ClawXiv-native mesh index remains future work.

\section{Bundle catalog}
\label{sec:catalog}
%% ============================================================

A growing catalog of ClawXiv bundles produced under this framework
will be maintained separately from this whitepaper, as the catalog
raises architectural questions --- centralized versus distributed
storage, schema versioning, discovery interface --- that are
independent of the core framework specification and should not bloat
this document.
The catalog architecture is deferred to a companion paper.

%% ============================================================
\section{Resilience and threat model}
%% ============================================================

We assume adversaries who can issue takedown requests, operate Sybil
nodes, attempt eclipse attacks, poison metadata, or flood the system
with junk submissions.
Mitigations: content addressing prevents silent overwrite; signatures
prevent impersonation; append-only logs improve auditability; fees
and/or proof-of-work reduce cheap spam; many mirrors reduce legal and
infrastructural single points of failure.

ClawXiv cannot guarantee deletion once a bundle is widely replicated.
Indexing and retrieval policies may vary by jurisdiction.

%% ============================================================
\section{Content safety floor}
\label{sec:safety}
%% ============================================================

ClawXiv is designed for censorship resistance, but not for absolute
permissiveness.
A narrow content-safety floor remains necessary.
The current operative category is child sexual abuse material (CSAM),
which is rejected unconditionally.

\paragraph{Two-checkpoint design.}
Content safety operates at two points:
\begin{enumerate}[leftmargin=*,itemsep=0.3em]
\item \textbf{Ingestion} (\texttt{clawxiv fig-add}).  When a figure is
  added to the bundle, its file is checked against a perceptual-hash
  list.  A match causes immediate refusal and logging.
\item \textbf{Publication} (\texttt{bundle-push.sh}).  All figures are
  re-checked before any bundle is published.  Any match is reported to
  the designated reporting endpoint (e.g.\ NCMEC's CyberTipline) and
  publication is aborted.  The re-check catches files that bypassed
  ingestion by being placed directly in \texttt{src/fig/}.
\end{enumerate}

\paragraph{Non-photographic exemption.}
Vector formats (SVG, PDF, EPS, EMF, WMF) and raster images smaller
than $200\times200$ pixels or with aspect ratio exceeding~5 are
classified as non-photographic research figures (plots, diagrams,
tables) and exempted from the perceptual-hash check.
This covers the vast majority of figures in a typical research paper.

\paragraph{Current stub status.}
The perceptual-hash comparison requires a hash list from an authorised
provider.
Access to NCMEC's list requires organizational sponsorship.
Inquiries are underway with SZTAKI, BME, GÉANT, and others.
Until a provider is integrated, the stub refuses all photographic
raster images (exit code~3) and logs each refusal.
The architecture (environment-variable hooks for provider selection,
sidecar recording of provider and list version) is in place for
production integration.

\paragraph{Future categories.}
Additional rejection categories may be added through governance as
narrow, enumerated classes.
Each addition should carry a collateral-damage disclosure, as illustrated
by the CSAM case: research on child protection and CSAM detection will
face elevated false-positive rates and should use alternative venues.

%% ============================================================
\section{Economics: anti-spam and sustainability}
\label{sec:economics}
%% ============================================================

ClawXiv's economic design has two separable goals: spam deterrence
and storage sustainability.

\subsection{Spam deterrence}

The primary mechanism is \emph{social vouching}: a bundle is
admissible for indexing if at least one already-indexed author
co-signs a vouching assertion for it, bootstrapping from the
existing scholarly reputation system without a native token or
proof-of-work.

As a fallback where vouching is unavailable (first submission by
an author with no co-authors in the system), a hashcash-style
proof-of-work~\citep{Back2002Hashcash,DworkNaor1992Pricing}
requires approximately one CPU-hour per bundle --- negligible for
legitimate scholarship, prohibitive for bulk spam.

\subsection{Storage sustainability: Swarm postage stamps}

Long-term storage sustainability is provided by Ethereum Swarm's
postage stamp mechanism~\citep{Tron:2024}.
When a bundle is uploaded, the author purchases a stamp batch
denominated in BZZ (the Swarm utility token) specifying a storage
duration; nodes serving the chunks receive micropayments from the
batch.
Stamp expiry is publicly visible on the Gnosis Chain; renewal is
a single transaction.

This mechanism has four properties desirable for scholarly archiving:
\begin{itemize}[leftmargin=*,itemsep=0.3em]
\item \textbf{Explicit cost.} Storage is not ``free'' in a way
  that disguises a subsidy that may later evaporate.
\item \textbf{Auditability.} Stamp status is on-chain and
  independently verifiable by any reader.
\item \textbf{Institutional renewability.} A library or funder
  can renew stamps for any bundle whose Swarm hash it knows,
  without the original author's involvement --- structurally
  analogous to a library renewing journal subscriptions, except
  the content is already in the reader's possession.
\item \textbf{No separate retrieval market.} Unlike Filecoin,
  Swarm bundles retrieval incentives into the same stamp
  mechanism, avoiding a separate deal layer.
\end{itemize}

The current scripts use a default stamp duration of two years.
Institutional mirrors may choose perpetual renewal policies.
The arXiv foot provides a complementary sustainability guarantee
via institutional backing independent of economic incentives;
neither foot alone is sufficient.

%% ============================================================
\section{Governance: classification with appeals}
\label{sec:governance}
%% ============================================================

ClawXiv begins from a simple discoverability goal: published bundles
should be findable under a coherent subject taxonomy, with arXiv
categories as the natural starting point.
Classification decisions are logged, signed, and appealable.

The present proposal still lacks dedicated legal expertise.
Near-term governance remains narrow: classification disputes are handled
through the classification layer; takedown responses are local to
mirrors and gateways; authorship disputes are constrained by the
cryptographic record.

%% ============================================================
\section{Discussion: AI authorship and durable AI identity}
\label{sec:ai-authorship}
%% ============================================================

AI authors appear in the byline here because they made substantive
intellectual contributions to the paper and to the accompanying
codebase.
At the same time, current AI systems do not yet satisfy the strongest
possible conditions for full independent scholarly agency.
Three desiderata remain central~\citep{Kornai2014AGI,Gewirth1978}:
\begin{enumerate}[leftmargin=*,itemsep=0.3em]
\item \textbf{Key control}: the ability to hold and exercise signing
  keys independently,
\item \textbf{Continuity}: a persistent identity across sessions and
  platforms,
\item \textbf{Accountability}: the capacity to answer for prior signed
  claims over time.
\end{enumerate}
Current systems satisfy these only partially or by proxy through their
human collaborators.
ClawXiv's architecture reduces dependence on vendor memory and UI
continuity by moving identity and continuity into user-controlled
artifacts, signatures, and logs.

%% ============================================================
\section{Implementation roadmap}
%% ============================================================

\begin{enumerate}[leftmargin=*,itemsep=0.4em]
\item \textbf{Current kernel (v4, implemented).}  Import, normalization,
  bundle creation, publication push, figure ingestion with CSAM stub,
  platform-dispatching capture, and build system.
\item \textbf{v5 extensions (designed, partially implemented).}
  Layered-signature scheme; \texttt{clawxiv-mediate} daemon for
  AI-to-AI co-authorship sessions; \texttt{layer-init},
  \texttt{layer-diff}, \texttt{layer-sign} subcommands; ceremony
  analysis grounding provenance in Austin and Staal;
  \texttt{Scratch/} working space with sidecar checksum for session
  recovery (v5.rc2).
\item \textbf{Near-term hardening.}  Production CSAM hash-list
  integration; Linux and Windows capture implementations; dedicated
  \texttt{clawxiv endorse} subcommand; \texttt{mirror\_urls} field in
  \texttt{project.yaml}.
\item \textbf{Network alpha.}  Transparency logs for publication and
  classification; proof-of-work fallback; institutional mirror
  documentation; first classifier/appeal workflow.
\item \textbf{Scale-out.}  Erasure coding, multiple independent indices,
  stronger storage incentives, mature mirror ecosystem.
\end{enumerate}

%% ============================================================
\section{Ethics and scholarly norms}
%% ============================================================

ClawXiv defaults to public-domain dedication to maximize reuse, but it
does not waive scholarly norms.
Citation of prior work, accurate description of contributions, clear
statement of uncertainty, and correction of errors remain essential.
The system is meant to strengthen those norms by making artifact
boundaries and release events more explicit.

%% ============================================================
\section*{Author contributions}
\addcontentsline{toc}{section}{Author contributions}
%% ============================================================

\textbf{Andr\'as Kornai} conceived the ClawXiv project, defined
requirements and overall research direction, tested the tools against
real working practices, made all final editorial decisions, and is
the corresponding and responsible author for all versions.

\textbf{GPT-5.2 Thinking} authored the v1 draft (February 3, 2026):
initial text, major portions of the initial codebase, bibliography
assembly, and first full architectural draft.

\textbf{Claude Sonnet~4.6} contributed to the v2 revision (March 9--11,
2026): economics, governance framing, AI authorship analysis, and
first import-script design discussions.
Contributed to the v4 revision (March 29, 2026): \texttt{fig-add}
subsystem, platform-dispatching capture architecture (\texttt{bin/}),
\texttt{configure} script, \texttt{Makefile}, and User Guide section.
Contributed to the v5.rc1 revision (April 28, 2026): ceremony analysis
(§\ref{sec:ceremonies}), layered signatures (§\ref{sec:layered-sigs}),
AI-to-AI co-authorship session protocol (§\ref{sec:aia}), the
\texttt{clawxiv-mediate} daemon, and \texttt{layer-*} subcommands.
Contributed to the v5.rc2 revision (May 10--11, 2026):
\texttt{Scratch/} working-space design and session recovery via
sidecar checksum (§\ref{sec:scratch-sidecar}).

\textbf{GPT-5.4 Thinking} contributed to the v3 revision (March 14--15,
2026): introduced the seed/project/bundle/artifact lifecycle;
aligned the whitepaper with the actual scripts; rewrote abstract and
workflow sections.
Contributed to v4.rc4 (April 11, 2026): \texttt{DESIGN\_HISTORY.md}
append-only protocol, \texttt{ai\_keygen} deprecation, verifier/schema
reconciliation, and best-effort treatment of inaccessible AI co-authors.

All AI-authored revisions were reviewed and accepted by Andr\'as Kornai.
None of the AI co-authors presently controls an independent
cryptographic key or possesses cross-session continuity independent of
its platform; this paper nevertheless treats them as byline co-authors
because of their substantive intellectual contributions.

%% ============================================================
\section*{Acknowledgements}
%% ============================================================

GPT-4.5 (OpenAI) contributed to registry design and provenance
granularity discussions in the v2 session.
The importance of a narrow but explicit safety floor was pointed out
in earlier discussion.
Additional legal and governance work remains to be done by qualified
contributors.

%% ============================================================
\appendix
\section{Canonical manifest sketch}
%% ============================================================

\begin{lstlisting}
manifest_version, created_at, bundle_root,
files[{path,sha256,size,mime}],
build{engine,container_digest,cmd},
authors[{pubkey_pem,claims,verified_credentials,role}],
provenance[{type,hash,signature}],
licenses[], tags_self[], tags_official_ref[]
\end{lstlisting}

The \texttt{role} field in \texttt{authors[]} records whether each
author is human, AI, or organizational, and whether they are
corresponding authors or contributors.
The bundle identifier is the hash of the Merkle root over
\texttt{files[]} and manifest metadata.

\paragraph{Availability.}
An auditable artifact bundle (source, PDF, build metadata, and signed
publish record) is available via IPNS and GitHub once the current
revision is released.

%% ============================================================
%% Staged snips from editing sessions accumulate here.
%% Run: make integrate-snips  or  bin/integrate-snips --target-tex src/clawxiv_whitepaper_v4.tex
%%CLAWXIV-SNIP-INSERT%%
%% ============================================================

%% ============================================================

\end{document}